\documentclass[twocolumn,floatfix,aps,superscriptaddress,prl]{revtex4-1}

\usepackage[utf8]{inputenc}
\usepackage{natbib}
\usepackage{graphicx}
\usepackage{xcolor}
\usepackage{bm}
\usepackage{mhchem}
\usepackage{physics}
\usepackage{amsfonts}
\usepackage[normalem]{ulem}
\usepackage{comment}

\usepackage{pdfpages}
\usepackage{pgffor}
\makeatletter
\AtBeginDocument{\let\LS@rot\@undefined}
\makeatother

\newcommand{\mc}[1]{{#1}}

\definecolor{cocol}{rgb}{0,0.6,0}

\definecolor{bluish}{rgb}{0,0.4,0.9}

\makeatletter
\g@addto@macro\normalsize{%
  \setlength\abovedisplayskip{5pt}
  \setlength\belowdisplayskip{5pt}
  \setlength\abovedisplayshortskip{5pt}
  \setlength\belowdisplayshortskip{5pt}
  \setlength{\belowcaptionskip}{-20pt}
}
\makeatother

\newcommand{\mbf}[1]{{\ensuremath{\mathbf{#1}}}  }
\newcommand{\mca}[1]{{\ensuremath{\mathcal{#1}}}  }

\newcommand{\br}{\mbf{r}}

\newcommand{\bG}{\mbf{G}}

\newcommand{\CX}{\mca{X}}

\newcommand{\inlineket}[1]{|{#1}\rangle}
\begin{document}

\title{On the Completeness of Atomic Structure Representations}

\author{Sergey N. Pozdnyakov}
\thanks{These two authors contributed equally}
\affiliation{Laboratory of Computational Science and Modelling, Institute of Materials, Ecole Polytechnique F\'ed\'erale de Lausanne, Lausanne 1015, Switzerland}
\author{Michael J.~Willatt}
\thanks{These two authors contributed equally}
\affiliation{Laboratory of Computational Science and Modelling, Institute of Materials, Ecole Polytechnique F\'ed\'erale de Lausanne, Lausanne 1015, Switzerland}
\author{Albert P. Bart\'ok}
\affiliation{Department of Physics and Warwick Centre for Predictive Modelling, School of Engineering, University of Warwick, Coventry CV4 7AL, United Kingdom}
\author{Christoph Ortner}
\email{c.ortner@warwick.ac.uk}
\affiliation{Mathematics Institute, University of Warwick, Coventry CV4 7AL, United Kingdom}
\author{G\'abor Cs\'anyi}
\email{gc121@cam.ac.uk}
\affiliation{Engineering Laboratory, University of Cambridge, Trumpington Street, Cambridge CB2 1PZ, United Kingdom}
\author{Michele Ceriotti}
\email{michele.ceriotti@epfl.ch}
 \affiliation{Laboratory of Computational Science and Modelling, Institute of Materials, Ecole Polytechnique F\'ed\'erale de Lausanne, Lausanne 1015, Switzerland}
\date{\today}

\begin{abstract}
Many-body descriptors are widely used to represent atomic environments in the construction of machine learned interatomic potentials and more broadly for fitting, classification and embedding tasks on atomic structures. It was generally believed that 3-body descriptors uniquely specify the environment of an atom, up to a rotation and permutation of like atoms. We produce several counterexamples to this belief, with the consequence that any classifier, regression or embedding model for atom-centred properties that uses 3 (or 4)-body features will incorrectly give identical results for different configurations. 
\mc{Writing global properties (such as total energies) as a sum of many atom-centred contributions mitigates, but does not eliminate, the impact of this fundamental deficiency -- explaining  the success of current ``machine-learning'' force fields. We anticipate the issues that will arise as the desired accuracy increases, and suggest potential solutions.}
\end{abstract}

\maketitle
Over the past decade tremendous progress has been made in the use of statistical regression to sidestep computationally demanding electronic structure calculations, and obtain ``machine-learning'' models
of materials and molecules, that use as inputs only the chemical nature and coordinates of the atoms \cite{behl-parr07prl, szla+14prb, bart+10prl, koba+17prm, lubb+18jcp, gast+17cs, glie+18prb, schu+17ncomm, gris+19acscs, wilk+19pnas}. A crucial driver of this progress has been the introduction of \emph{representations} of atomic structures:  A property associated with the $i$-th atom can be written as 
$F_i = \mathcal{F}\left(\CX_i\right)$, 
where $\CX_i=\{\br_{ij}\}_{j \neq i}$ describes the neighbour  environment of the $i$-th atom. 
To preserve symmetries of the target property, the representation of $\CX_i$ should be equivariant~\cite{glie+17prb,gris+18prl} (often simply invariant~\cite{behl-parr07prl,bart+10prl,bart+13prb,sade+13jcp,zhu+16jcp}) with respect to translations,  rotations, labelling of identical atoms, and often also reflections.
Most of the invariant representations  \cite{behl-parr07prl, bart+10prl,bart+13prb, Chmiela2016, vonl18ac} can be seen as projections onto different bases of  many-body correlation functions \cite{will+19jcp}. 
To stress that our results apply equally to all these frameworks, we use the abstract notation $|\CX_i^{(\nu)}\rangle$ to indicate the $(\nu+1)$-body correlation, which is centered on the $i$-th atom \cite{will+19jcp}.
For instance, the 2-body correlation $|\CX^{(1)}\rangle$ corresponds to the histogram of interatomic distances $r_{ij}$ -- equivalent to the radial distribution function or the 2-body symmetry functions, $G_2$, of Ref.~\cite{behl-parr07prl}.
The 3-body correlation $|\CX^{(2)}\rangle$  is equivalent to the histogram of triangles, represented by the 3-tuples $\left(r_{ij},r_{ij'},\omega_{ijj'}=\hat{\bf r}_{ij} \cdot \hat{\bf r}_{ij'}\right)$
 -- and to the power spectrum~\cite{bart+10prl}, or to the 3-body symmetry functions, $G_3$~\cite{behl-parr07prl}. 
Linear regression based on these features is equivalent to a body-ordered expansion of the target property\cite{glie+18prb,thom+15jcp,will+19jcp,drau19prb,VanderOord2019,Shapeev}. Given that computing higher-order terms is increasingly costly, the representation is typically truncated at 3 or 4 body correlations.

\begin{figure}[t]
    \centering
    \includegraphics[width=0.95\columnwidth]{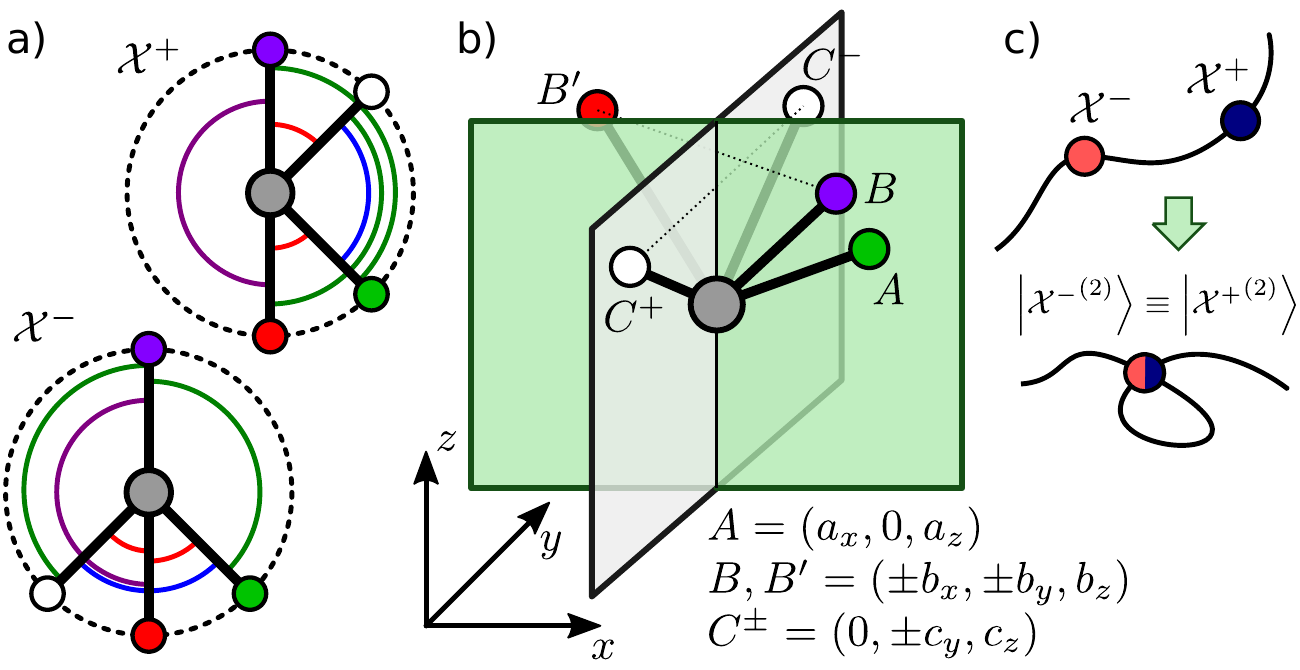}
\caption{ (a) Two structures with the same histogram of triangles; (angles: 45$^\circ$, 45$^\circ$, 90$^\circ$, 135$^\circ$, 135$^\circ$, 180$^\circ$) (b) A manifold of degenerate pairs  of environments: In addition to three points $A, B, B'$ a fourth point $C^{+}$ or $C^{-}$ is added leading to two degenerate environments, $\CX^+$ and $\CX^-$. (c) Degeneracies induce a transformation of feature space so that structures that should be far apart are brought close together.} 
\label{fig:sergei-michael}
\end{figure}

Employing \emph{non-linear} functions of low-order invariants, e.g. $F_i=\tilde{\mathcal{F}}(|\CX_i^{(2)}\rangle)$, incorporates information on higher-order correlations, and there is a widespread belief in the community, \mc{supported by numerical evidence~\cite{bart+13prb}} that the 3-body correlation unequivocally identifies an atomic environment. This {\em completeness} (injectivity) of the structure-representation map would guarantee that any atom-centered property can be described by $\tilde{\mathcal{F}}$, which extends to any atom-centered decomposition of extensive properties, such as the total energy\cite{glie+18prb}.
In this Letter, we present several counterexamples to this widely-held belief,  discuss the implications for machine learning atomistic properties,  and suggest directions towards the construction of complete representations. 

Figure~\ref{fig:sergei-michael}a  exhibits a simple example of a pair of environments, $\CX^+$ and $\CX^-$, with four neighbouring atoms of the same species positioned on a circle around the central atom.
The two structures cannot be superimposed by rotations and mirror symmetry, but they have the same list of distances and angles and hence cannot be distinguished by their 3-body correlations.
To elucidate this example, and more generally understand the difficulty of reconstructing an atomic environment from a body order representations, consider the Gram matrix $G_{jj'} = \br_{ij}\cdot \br_{ij'}$,
which contains sufficient information to reconstruct a configuration up to an arbitrary rotation or reflection.
If all the distances $r_{ij}$, or the chemical identity of the neighbors, are distinct, one can unequivocally assign distances and angles to a specific atom, and reconstruct the Gram matrix from the unordered list $\{(r_{ij},r_{ij'},\omega_{ijj'})\}$.
If some of the distances are the same, however, it becomes possible to swap some entries of $\bG$, yielding two or more degenerate environments that are different, but have the same 3-body invariants. %

\begin{figure}[!tbph]
    \centering
    \includegraphics[width=0.95\columnwidth]{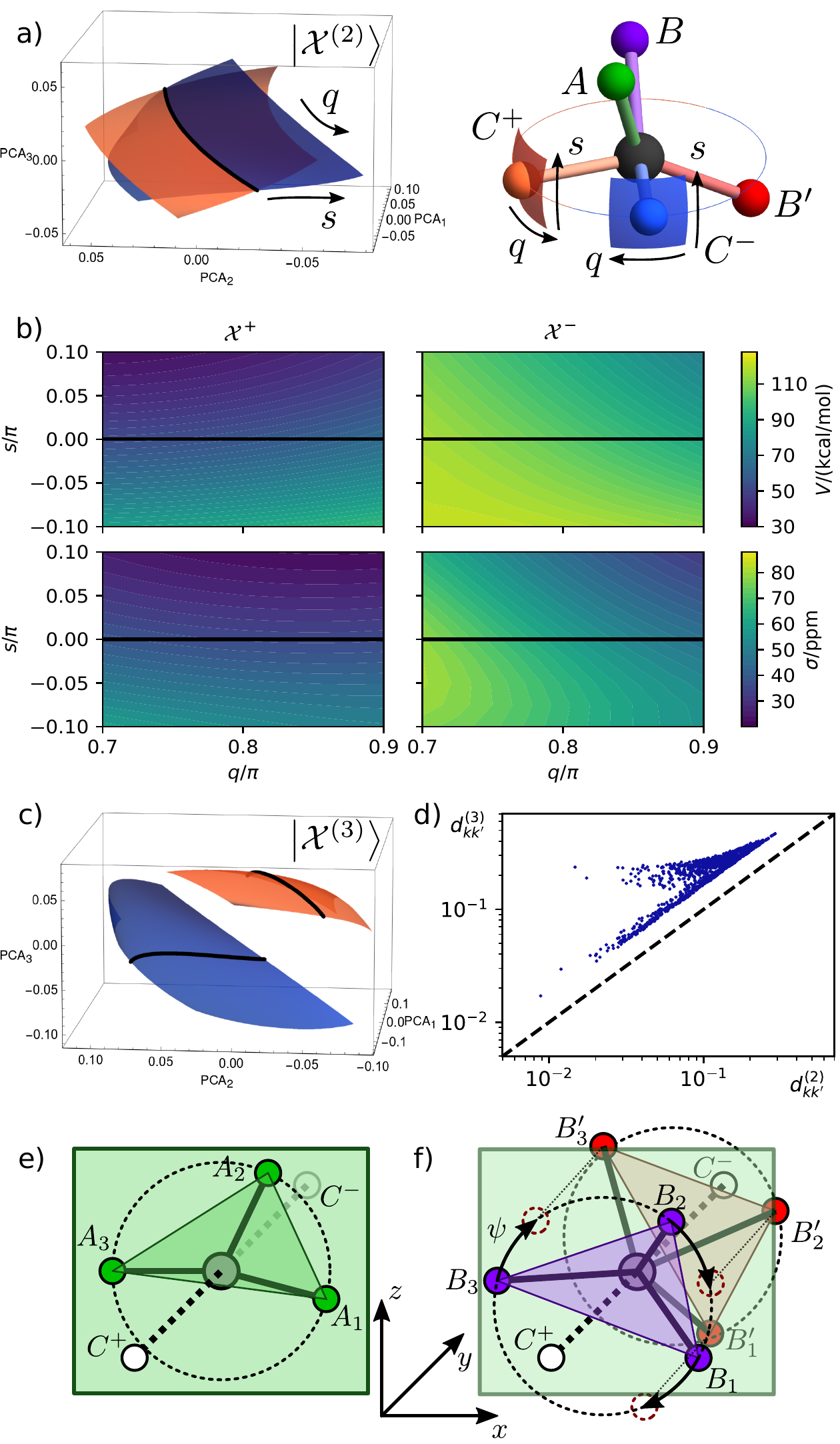}
\caption{
(a) PCA projection of $|{\CX}^{(2)+}\rangle$ and $|{\CX}^{(2)-}\rangle$ for a continuous manifold of \ce{CH4} environments $\CX^+$ and $\CX^-$, parameterised by $q$ (that moves along the degenerate set, represented by a black line) and $s$ (that breaks the degeneracy). 
(b) Energy (top) and $^{13}$C chemical shieldings (bottom) of a \ce{CH4} molecule that follows such manifolds; the zero of the two quantities is set to the values for the ideal geometry. (c) PCA projection of the bispectrum $\inlineket{\CX^{(3)}}$ space manifold. (d) Correlation plot of the distances between two points $k$ and $k'$ along both manifolds, computed based on the power spectrum ($d_{kk'}^{(2)}$) or the bispectrum ($d_{kk'}^{(3)}$).
(e) Construction of a pair of environments that are mirror images but share identical chiral $\inlineket{\CX^{(3)}}$ features. $A$ points lie in the $xz$ plane, along a circle centred on the origin. $C^\pm$ points lie along the $y$ axis, symmetric about the origin. (f) a pair of inequivalent structures with the same chiral $\inlineket{\CX^{(3)}}$ features.  $B$ and $B'$ points lie on circles centred on the origin, and shifted by the same amount above and below the $xz$ plane. One of the sets of points is twisted around $y$ by an angle $\psi$. }
    \label{fig:tetra}
\end{figure}

As shown in Fig.~\ref{fig:sergei-michael}b, one can generalize the construction to obtain a manifold of degenerate environment pairs  parameterised by 7 continuous variables. The total dimensionality of the configuration space of 4 neighbours is $4\times 3 - 3 = 9$. Thus, the degenerate manifold has a dimension of 7 and a codimension of 2.  When going from the $+$ to the $-$ structure in the pair, the elements of the Gram matrix between $C$-type and $B$-type points are swapped, leading to non-equivalent structures that have the same 3-body description. 
This construction can be extended by adding  further $A$ or $C$-type points (increasing the codimension of the degenerate manifold by one) or pairs of $B$-type points (each pair increasing the codimension by three).
Other counterexamples can be found, involving triplets of degenerate structures (see SI). 
Tight bounds on the codimension of degenerate manifolds and on the multiplicity of degenerate structures is a key aspect in understanding the success of incomplete environment descriptors, but is beyond the scope of the present work. However, the example of Fig.~\ref{fig:sergei-michael}b is sharp in the sense that (i) for three or fewer neighbours the 3-body correlation suffices to reconstruct the enviroment and (ii) for four or more neighbours one can construct a manifold of co-dimension 2 which must contain all degenerate environments. These results, which build on those in Ref.~\cite{Boutin2004}, are detailed in the SI. It is unclear to us whether
the increase of the co-dimension when neighbors are added in the example of Fig.~\ref{fig:sergei-michael}b is specific to our construction, or reflects a general result.

Following the procedure in Fig.~\ref{fig:sergei-michael}b, one can produce a pair of degenerate tetrahedral environments, that we label $\CX^+$ and $\CX^-$, corresponding to a \ce{CH4} molecule.
Figure~\ref{fig:tetra}a shows a portion of the two manifolds (blue and red surfaces, parameterised by two variables $q$ and $s$) built as a principal component projection of the power spectrum space (details given in the SI).
Structures within the two surfaces correspond to configurations that are different from each other, but those along the black line (corresponding to $s=0$) have identical 2- and 3-body invariants, which therefore cannot distinguish $\CX^+$ and $\CX^-$, and the two manifolds intersect each other. As shown in  Fig.~\ref{fig:tetra}b, however, both atom-centred properties such as the $^{13}$C NMR chemical shift, and extensive properties such as molecular energy, are very different as they cannot be described fully by 3-body correlations around the central atom.
\mc{Higher body-order features can differentiate between $\CX^+$ and $\CX^-$. 
As shown in Fig.~\ref{fig:tetra}c,  the feature-space degeneracy is lifted by the 4-body correlation (bispectrum), $|\CX^{(3)}\rangle$, which corresponds to the unordered list of tetrahedra formed by the central atom and three of its neighbors. 
The presence of a degeneracy can be revealed by comparing environment distances $d^{(2)}, d^{(3)}$ computed, respectively, from power spectrum coordinates $|\CX^{(2)}\rangle$ and bispectrum coordinates $|\CX^{(3)}\rangle$. One then observes that pairs of environments that are close in $d^{(2)}$ remain well separated by $d^{(3)}$ (Fig.~\ref{fig:tetra}d).
However, the bispectrum is not complete either. While it does differentiate between the tetrahedral \ce{CH4} environments, one can build pairs of environments that have the same 4-body correlations without being superimposable by proper (Fig.~\ref{fig:tetra}e) or improper (Fig.~\ref{fig:tetra}f) rotations.
Note that the environments in Fig.~\ref{fig:tetra}e are chiral (mirror) images of each other, but the bispectrum does not distinguish them because the tetrahedra it is composed of are {\em not} chiral.~\footnote{All body-order correlations above $\nu=3$, when defined as an average over proper rotations, are sensitive to chirality and can differentiate between enantiomers. When learning non-chiral properties, such as the energy,  one can average over inversion. Unless otherwise specified, in this work {$|\CX^{(3)}\rangle$} indicates the non-chiral version.}
}

\begin{figure}
    \centering
    \includegraphics[width=1.0\columnwidth]{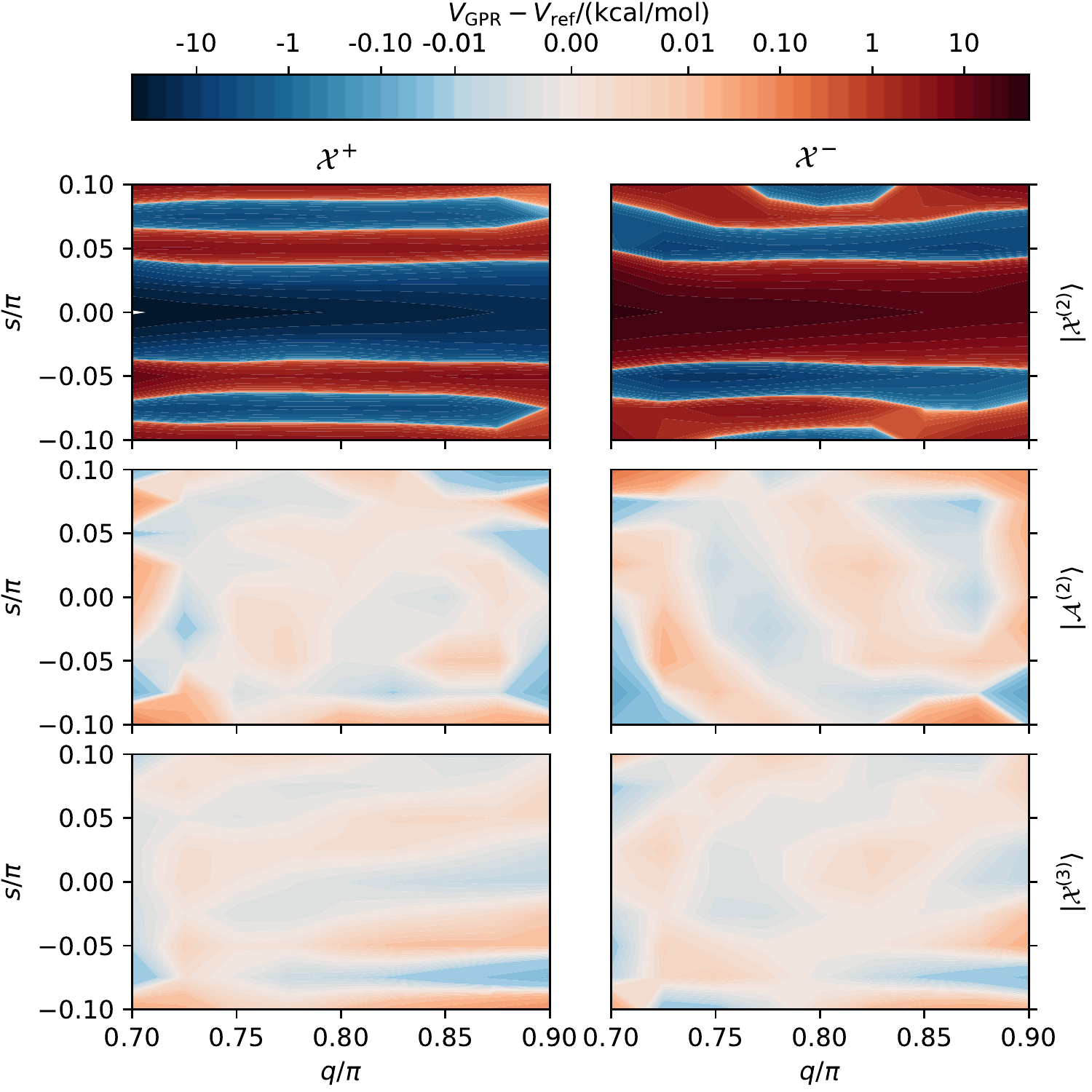}
\caption{Error in the prediction of the molecular energy for \ce{CH4} configurations along the manifold depicted in Fig.~\ref{fig:tetra}c and d, using a GPR model based on a non-linear kernel built on the \ce{C}-centred SOAP power spectrum (top, RMSE: 12kcal/mol), a combination of \ce{C} and \ce{H}-centred power spectra (middle, RMSE: 0.027 kcal/mol), and the \ce{C}-centred bispectrum (bottom, RMSE: 0.011 kcal/mol).}
\label{fig:models}
\end{figure}

A Gaussian process regression model based on a non-linear kernel built on the SOAP power spectrum (equivalent to the 3-body correlation, $|\CX^{(2)}\rangle$, see SI) results in large errors, not just along the $s=0$ line of degeneracy, but also for structures that are not exactly indistinguishable according to the power spectrum (top panels in Fig.~\ref{fig:models}). This underscores the fact that the existence of manifolds of degenerate structures introduces a distortion of the feature space (Fig.~\ref{fig:sergei-michael}c), and hinders the ability to perform  regression regardless of whether strictly degenerate pairs are included in the training. 
Because they are ultimately based on the same unordered sets of triangles, Behler-Parrinello ``atom-centered symmetry functions''\cite{behl-parr07prl}, the FCHL descriptors of von Lilienfeld and coworkers\cite{fabe+18jcp}, the MBTR descriptor of Rupp \cite{Huo2017}, and the smooth version of the DeepMD framework~\cite{DeepMD2018} will also suffer from the same problem. 
\mc{
The fact that a large manifold of \ce{CH4} environments is un-learnable using 2- and 3-body features is a shortcoming, that fundamentally limits the reliability of machine-learned models of atom-centred properties based on these descriptors.}

\begin{figure}
\centering
\includegraphics[width=1.0\linewidth]{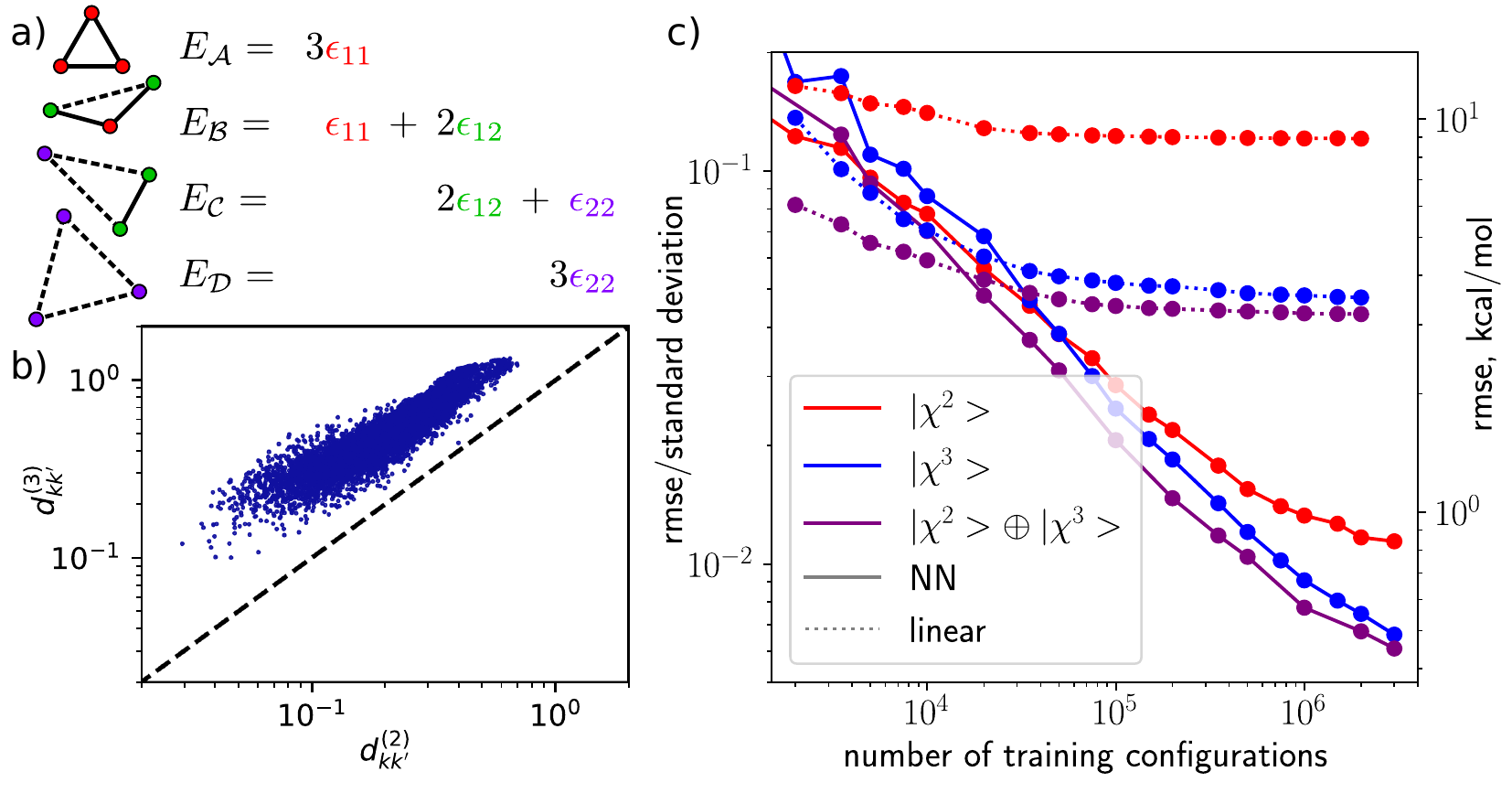}
\caption{(a) Four configurations distinguishable by their atom-centered 2-body histograms. Only three different site energies occur in these configurations, hence fitting four total energies leads to overdetermined regression. (b) Correlation plot of powerspectrum and bispectrum distances between \ce{C} environments in a database of random \ce{CH4} configurations. (c) Learning curves for the atomization energy of random \ce{CH4} configurations.  }
\label{fig:degenerate-model}
\end{figure}

\mc{
When learning the decomposition of a \emph{global} property, such as the total energy, one can hope to lift the degeneracy by using features centred on other atoms in the structure. For the construction in Fig.~\ref{fig:sergei-michael}b, there is always at least one atom outside the bisecting $A$ plane that breaks the indistinguishability of $\CX^+$ and $\CX^-$.
Indeed, a model that combines \ce{C} and \ce{H}-centred non-linear kernels can approximate the molecular energy to excellent accuracy, also along the degenerate manifold (see Fig.~\ref{fig:models}, middle panels).
In general, however, one {\em cannot} rely on such a mechanism. For the sake of simplicity, we demonstrate this for the case of 2-body descriptors $\inlineket{\CX^{(1)}}$. It is well-known that the list of distances from the centre of an environment, or even the list of distances in a structure~\cite{Boutin2004}, are not complete representations. 
It has, however, been speculated~\cite{vonl+15ijqc} that simultaneous knowledge of all atom-centred lists of distances in a structure would provide a complete representation of the configuration, and that one could use this representation to predict arbitrary potentials using an additive model based on non-linear functions of $\ket{\CX^{(1)}}$. Both conjectures are false. We present a counter-example to the first conjecture in the SI. 
The counterexample to the second statement, cf. Figure~\ref{fig:degenerate-model}a, is far more concerning though: even if, in a training set, all configurations can be uniquely identified by the collection of the atom-centered 2-body histograms, it does not follow that a total energy represented in terms of these histograms can be learned. 
}

\mc{
The breakdown of the purely 2-body models in these limiting cases has practical implications, as they translate into instability and data inefficiency in real-life scenarios -- which is the ultimate reason why models based on purely radial information have been superseded by those incorporating 3-body features.
Proving the existence of similar counterexamples for the learning of global properties using $\inlineket{\CX^{(2)}}$ is more challenging. It is possible, however, to numerically demonstrate how a model based on 3-body features suffers from a degradation of learning efficiency, provided that one pushes it to sufficiently high  accuracy. Figure~\ref{fig:models}b,c show results for a data set of about 3 million \ce{CH4} configurations obtained by randomly distributing the atoms and discarding structures with too close contacts (details in the SI). The distance-distance correlations (panel b) show that there are configurations that approach the degenerate manifolds, but there are no fully-degenerate pairs.
We then built an additive model that includes contributions from both the \ce{C} and the \ce{H} atoms, converging the discretization of $\ket{\CX^{(2)}}$ and using a neural network to ensure maximal flexibility in the feature-property mapping.
The learning curves (Fig.~\ref{fig:models}c) exhibit clear signs of saturation, indicating that even though each pair of environments (and therefore structures) in the data set can be distinguished based on $\ket{\CX^{(2)}}$, the presence of near-degeneracies affects the stability and efficiency of the regression. 
}

%
%
%
\mc{
Using the higher-body order features to differentiate between $\CX^+$ and $\CX^-$ does indeed lead to a more efficient model (Fig.~\ref{fig:models}, bottom panel), that predicts the energy along the degenerate manifold with an error that is roughly a third of that obtained by a multi-center, power-spectrum-based model. Substantial improvements are also seen for the random  \ce{CH4} configurations. A NN based on $\inlineket{\CX^{(3)}}$ reduces the full-train-set error by 40\%{}, down to $\approx 0.5$ kcal/mol. Similar to what was observed for $\ket{\CX^{(2)}}$-based models that combine multiple cutoff distances~\cite{will+18pccp}, there is a data/complexity tradeoff. For small training set sizes a simpler powerspectrum model can outperform one based on the bispectrum, and linear regression outperforms a deep neural network. 
The best balance between data efficiency, computational cost and ultimate accuracy might involve a combination of different kinds of features, as demonstrated by the hybrid model in Fig.~\ref{fig:degenerate-model}. Approaches such as the moment tensor potentials~\cite{Shapeev}, permutationally invariant polynomials\cite{Braams:2009eb,vanderOord:2020gx} and the atomic cluster expansion~\cite{drau19prb} allow, if necessary, to further resolve degeneracies by including arbitrary body-orders of correlation.
We show in the SI that similar considerations apply also to a database of bulk silicon structures~\cite{Bartok2018a}. The cutoff distance, however, complicates the picture, because the number of neighbors included in the environments influences the proximity of structures to the degenerate manifold, and because the model accuracy is also affected by the truncation of long-range interactions~\cite{gris-ceri19jcp}. }
Descriptors such as eigenspectra of matrices constructed from the atomic configuration (distance matrix, Laplacian, orbital overlap, etc.)\cite{Goedecker:2016ky} also contain information on high body order correlations, and as such are not expected to be degenerate for the present examples. Their completeness properties are not understood at present.

Overall, the results we have shown indicate that despite the remarkable success of ML models that describe atomic structures in terms of $n$-body correlations features, there is still work to do to understand fully how the configuration space of a set of atoms is mapped onto symmetry-adapted representations. 
The problem is to construct a representation %
which is (i)~complete; (ii)~smooth with smooth inverse; (iii)~and invariant under isometries and permutations. 
An obvious, but ineffective, solution is to use %
the union of {\em all} $n$-point correlations \cite{Shapeev,drau19prb}.
Pragmatically, one can proceed as we do here for the \ce{CH4} dataset,  increasing the correlation order until all configurations in a given training set are distinguishable, possibly reducing the cost of computing high-order features using a sparsification procedure along the lines of \cite{imba+18jcp}. %
It is, however, desirable to know {\it a priori} which features are required to guarantee (i--iii). 
For example, we may ask whether there is a fixed finite $\bar{n}$ such that all higher-order $n$-points correlations can be recovered from the $\bar{n}$-point correlation.
There are at least two perspectives from which to pursue questions of this kind: signal processing and invariant theory.

In the signal processing literature it has long been known that the power spectrum  is insufficient to reconstruct {\em most} signals, while the bi-spectrum  uniquely identifies translation-invariant and compact signals~\cite{Yellott1992-gh,Kondor2018,kakarala2012}. On the other hand, Ref.~\cite{Yellott1992-gh} provides a range of elementary examples establishing that no correlation order suffices to reconstruct all periodic signals.
Nevertheless, stable bispectrum inversion has been shown to work well in practice due to the fact the {\em most} signals can be reconstructed from it; see e.g.~\cite{Bendory2018-uh,Bandeira2017-le} and references therein. 
These results have a striking parallel to our own observations regarding the reconstruction of an atomic environment and in particular suggest that {\em in theory} no $\bar{n}$-point correlation may suffice to reconstruct the environment. %

Still, since atomic environments can be thought of as a very restrictive class of signals, the invariant theory perspective may shed additional light on our questions. The perspective of Boutin and Kemper~\cite{Boutin2004} appears to be particularly useful, establishing conditions under which a points cloud can be reconstructed from the histogram of distances.
The problem we tackle here is closely related: degeneracy of two centred environments with respect to $n$-body correlations implies degeneracy of the point clouds consisting of the neighbors with respect to $n-1$ body correlations. For example, Fig.~\ref{fig:sergei-michael}a, implies that the length-histogram of the neighbours lying on the circle are degenerate (indeed, this is the example given in Fig.~4 in Ref.\cite{Boutin2004} and in Fig 2 of \cite{vonl+15ijqc}). Similarly, Fig~\ref{fig:tetra}f, shows  {\em environments} that are degenerate with respect to the 4-body correlation (tetrahedron histograms) are also degenerate with respect to the 3-body correlations (triangle histograms) of the {\em entire} structure. 
A similar approach may therefore help determine tight bounds on the codimension of the degenerate manifold although, as far as we are aware, there are no rigorous results in this direction.
\mc{
The problem of formulating a complete feature map is of fundamental importance -- particularly when considering the use for generative models that require inverting  the relation between a representation and the underlying structure -- and has practical implications, particularly when one wants to achieve high accuracy with the minimum amount of data. 
The presence of many neighbors or of different species (that provide distinct ``labels'' to associate groups of distances and angles to specific atoms), and the possibility of using representations centred on nearby atoms to lift the degeneracy of environments reduces the detrimental effects of the lack of uniqueness of the power spectrum when learning extensive properties such as the energy. %
We show, however, that the learning rate of this kind of models reduces dramatically in the high accuracy regime, revealing the limitations of a description based on 3-body features. Diagnostic tools such as the joint distance histogram that we introduce here can help identify problematic parts of datasets, give more confidence in the reliability of simple-to-compute low-order invariants, and guide the choice of a small number of higher-order features to improve the accuracy and efficiency of models.
}

\begin{acknowledgments}
MJW, SP and MC acknowledge funding by the Swiss National Science Foundation (Project No. 200021-182057). CO acknowledges funding by the Leverhulme trust,  RPG-2017-191.
\end{acknowledgments}

\clearpage
\foreach \x in {1,2,3,4,5,6,7,8,9,10,11}
{%
\clearpage


\section*{Supplementary material}

# On the Completeness of Atomic Structure Representations
## *Supplementary material*

## 1 Power spectrum

### 1.1 Moving points

Take two points $\mathbf{r}_+$ and $\mathbf{r}_-$ which share the same distance to the origin, $r_+ = r_-$, and are not parallel or antiparallel. It will prove useful to construct a complete orthonormal basis from the two points as follows,

$$\hat{\mathbf{i}} = \frac{\mathbf{r}_+ + \mathbf{r}_-}{|\mathbf{r}_+ + \mathbf{r}_-|} \tag{1}$$

$$\hat{\mathbf{j}} = \frac{\mathbf{r}_+ - \mathbf{r}_-}{|\mathbf{r}_+ - \mathbf{r}_-|} \tag{2}$$

$$\hat{\mathbf{k}} = \frac{\mathbf{r}_+ \times \mathbf{r}_-}{|\mathbf{r}_+ \times \mathbf{r}_-|}. \tag{3}$$

Note that $\hat{\mathbf{i}}$ and $\hat{\mathbf{j}}$ are orthogonal because $r_+ = r_-$.

Our goal is to introduce more points in such a way that all distance-angle triplets formed by pairs of points and the origin are the same in two configurations $(+)$ and $(-)$, formed by removing either $\mathbf{r}_-$ or $\mathbf{r}_+$. This will ensure the origin-centred power spectra of $(+)$ and $(-)$ are equal too. Any point we add falls into one of two classes.

### 1.2 Points in the bisecting plane

The bisecting plane is spanned by $\hat{\mathbf{i}}$ and $\hat{\mathbf{k}}$. Any point with no component along $\hat{\mathbf{j}}$ lies in this plane and makes the same angle with $\mathbf{r}_+$ and $\mathbf{r}_-$. Adding points in this plane contributes exactly the same distance-angle triplets to $(+)$ and $(-)$ and does not break indistinguishability. But if one only adds points in this way, the bisecting plane is a plane of symmetry before $\mathbf{r}_+$ or $\mathbf{r}_-$ is removed, and $(+)$ and $(-)$ are therefore mirror images of each other. We must therefore add at least one point outside the bisecting plane.

### 1.3 Points outside the bisecting plane

Any point outside the bisecting plane has a non-zero $\hat{\mathbf{j}}$ component,

$$\mathbf{r} = a\hat{\mathbf{i}} + b\hat{\mathbf{j}} + c\hat{\mathbf{k}}. \tag{4}$$

Such a point makes different angles, $\theta_+$ and $\theta_-$, with $\mathbf{r}_+$ and $\mathbf{r}_-$ and contributes different distance-angle triplets to $(+)$ and $(-)$, breaking indistinguishability. To recover indistinguishability, we must therefore add another point outside the bisecting plane,

$$\mathbf{r}' = d\hat{\mathbf{i}} + e\hat{\mathbf{j}} + f\hat{\mathbf{k}}. \tag{5}$$

If $r \neq r'$ indistinguishability is not recovered, so we must have

$$a^2 + b^2 + c^2 = d^2 + e^2 + f^2. \tag{6}$$

1

Since only (+) includes $\theta_+$ and only (−) includes $\theta_-$, the only way to recover indistinguishability is by introducing $\theta_-$ into (+) and $\theta_+$ into (−) with the new point $\mathbf{r}'$. This in combination with Eq. (6) and $r_+ = r_-$ is equivalent to the following constraints,

$$\mathbf{r} \cdot \mathbf{r}_+ = \mathbf{r}' \cdot \mathbf{r}_- \tag{7}$$

$$\mathbf{r} \cdot \mathbf{r}_- = \mathbf{r}' \cdot \mathbf{r}_+. \tag{8}$$

The sum and difference of these equations gives

$$\mathbf{r} \cdot \hat{\mathbf{i}} = \mathbf{r}' \cdot \hat{\mathbf{i}} \tag{9}$$

$$\mathbf{r} \cdot \hat{\mathbf{j}} = -\mathbf{r}' \cdot \hat{\mathbf{j}}, \tag{10}$$

which fixes $\mathbf{r}'$ to be

$$\mathbf{r}' = a\hat{\mathbf{i}} - b\hat{\mathbf{j}} + f\hat{\mathbf{k}}. \tag{11}$$

The length constraint Eq. (6) fixes the remaining component up to its sign,

$$\mathbf{r}' = a\hat{\mathbf{i}} - b\hat{\mathbf{j}} \pm c\hat{\mathbf{k}}. \tag{12}$$

We choose the solution with the minus sign because otherwise $\mathbf{r}$ and $\mathbf{r}'$ are reflections of each other through the bisecting plane and (+) and (−) are mirror images, which is precisely what set out to avoid.

## 1.4 In combination

By adding points in the two ways just described, one constructs (+) and (−) configurations that are indistinguishable through their origin-centred power spectra. Since there must be at least one point outside the bisecting plane to avoid mirror images, and such points can only be present in pairs, the simplest (+) and (−) configurations comprise four points excluding the centre: one point in the bisecting plane, two outside it and $\mathbf{r}_+$ or $\mathbf{r}_-$.

A transparent parameterisation is obtained by aligning $\hat{\mathbf{i}}$, $\hat{\mathbf{j}}$ and $\hat{\mathbf{k}}$ with the $x$, $y$ and $z$ axes. The four-particle case is depicted in Fig. 1 of the main text, with $\hat{\mathbf{i}}$ aligned along the $z$ axis, $\hat{\mathbf{j}}$ along the $y$ axis and $\hat{\mathbf{k}}$ along the $x$ axis,

$$\mathbf{r}_+ = a\hat{\mathbf{y}} + b\hat{\mathbf{z}} \tag{13}$$

$$\mathbf{r}_- = -a\hat{\mathbf{y}} + b\hat{\mathbf{z}} \tag{14}$$

$$\mathbf{r}_1 = c\hat{\mathbf{x}} + d\hat{\mathbf{z}} \tag{15}$$

$$\mathbf{r}_2 = e\hat{\mathbf{x}} + f\hat{\mathbf{y}} + g\hat{\mathbf{z}} \tag{16}$$

$$\mathbf{r}_3 = -e\hat{\mathbf{x}} - f\hat{\mathbf{y}} + g\hat{\mathbf{z}}. \tag{17}$$

There are seven free parameters, $a, b, c, d, e, f, g$, leaving a codimension of two. Adding an extra point in the bisecting plane increases the codimension by one, and adding an extra pair of points outside the bisecting plane increases the codimension by three. In spherical polar coordinates we have

$$\mathbf{r}_+ = r\left[\cos(q)\hat{\mathbf{y}} + \sin(q)\hat{\mathbf{z}}\right] \tag{18}$$

$$\mathbf{r}_- = r\left[-\cos(q)\hat{\mathbf{y}} + \sin(q)\hat{\mathbf{z}}\right] \tag{19}$$

$$\mathbf{r}_1 = r'\left[\cos(\alpha)\hat{\mathbf{x}} + \sin(\alpha)\hat{\mathbf{z}}\right] \tag{20}$$

$$\mathbf{r}_2 = r''\left[\sin(\theta)\cos(\phi)\hat{\mathbf{x}} + \sin(\theta)\sin(\phi)\hat{\mathbf{y}} + \cos(\theta)\hat{\mathbf{z}}\right] \tag{21}$$

$$\mathbf{r}_3 = r''\left[-\sin(\theta)\cos(\phi)\hat{\mathbf{x}} - \sin(\theta)\sin(\phi)\hat{\mathbf{y}} + \cos(\theta)\hat{\mathbf{z}}\right], \tag{22}$$

which is a special case of

$$\mathbf{r}_+ = r\left[-\sin(s)\hat{\mathbf{x}} + \cos(s)\cos(q)\hat{\mathbf{y}} + \cos(s)\sin(q)\hat{\mathbf{z}}\right] \tag{23}$$

$$\mathbf{r}_- = r\left[-\sin(s)\hat{\mathbf{x}} - \cos(s)\cos(q)\hat{\mathbf{y}} + \cos(s)\sin(q)\hat{\mathbf{z}}\right]. \tag{24}$$

Any variation of $s$ away from $s = 0$ breaks the indistinguishability of (+) and (−). The variation of $q$ and $s$ with every other parameter fixed is used to explore the (+) and (−) property surfaces in the main text (Fig. 2 onwards).

## 1.5 Moving the centre

For any point $\mathbf{r}$ outside the bisecting plane we have

$$|\mathbf{r}_+ - \mathbf{r}| \neq |\mathbf{r}_- - \mathbf{r}|, \tag{25}$$

because $r_+ = r_-$. By centring on the point $\mathbf{r}$ indistinguishability of the atom configurations is broken since the distance-angle triplet $(|\mathbf{r}_+ - \mathbf{r}|, |\mathbf{r}_+ - \mathbf{r}|, 0)$ appears once more in $(+)$ than in $(-)$. As mentioned previously, there must be at least one atom outside the bisecting plane for the atom configurations $(+)$ and $(-)$ not to be mirror images of each other, so distinguishability of the atom configurations can always be recovered by centring the power spectrum on such an atom.

## 1.6 More moving points

Throughout the discussion above we supposed the configurations $(+)$ and $(-)$ differ only in the position of two points, $\mathbf{r}_+$ and $\mathbf{r}_-$. We added points to both $(+)$ and $(-)$ satisfying constraints involving only distances of points to the origin and dot products with unit vectors $\hat{\mathbf{i}}, \hat{\mathbf{j}}, \hat{\mathbf{k}}$ generated by $\mathbf{r}_+$ and $\mathbf{r}_-$.

Suppose instead we have multiple points like $\mathbf{r}_+$ which feature only in $(+)$ and $\mathbf{r}_-$ which feature only in $(-)$. There must of course be the same number of points in $(+)$ and $(-)$ for the configurations to be indistinguishable, so we can group together the $\mathbf{r}_+$ and $\mathbf{r}_-$ points into pairs. Suppose we choose these pairs in a special way, such each point in a pair is the same distance to the origin and every pair generates the same unit vectors through Eqs. (1-3). Then the $k^{\text{th}}$ pair has the following form,

$$\mathbf{r}_+^k = q_k \hat{\mathbf{i}} + \Delta_k \hat{\mathbf{j}} \tag{26}$$

$$\mathbf{r}_-^k = q_k \hat{\mathbf{i}} - \Delta_k \hat{\mathbf{j}}. \tag{27}$$

Clearly,

$$\mathbf{r}_-^k \cdot \mathbf{r}_-^l = \mathbf{r}_+^k \cdot \mathbf{r}_+^l, \tag{28}$$

which shows that every distance-angle triplet in $(+)$ involving the centre and two $\mathbf{r}_+$ points is the same as a distance-angle triplet in $(-)$ involving the centre and two $\mathbf{r}_-$ points. Furthermore, since $\hat{\mathbf{i}}, \hat{\mathbf{j}}, \hat{\mathbf{k}}$ are common to all pairs, the constraints for adding the same points to $(+)$ and $(-)$ are identical to before. In other words, adding a point $\mathbf{r}'_+$ to $(+)$ and $\mathbf{r}'_-$ to $(-)$ preserves indistinguishability of $(+)$ and $(-)$ if these points have the form of Eqs. (26-27). Adding such a pair increases the codimension by one.

## 1.7 High-multiplicity degeneracies

The construction discussed in the previous sections generates pairs of structures with the same power spectrum. It is also possible, however, to construct sets of more than two degenerate structures. An

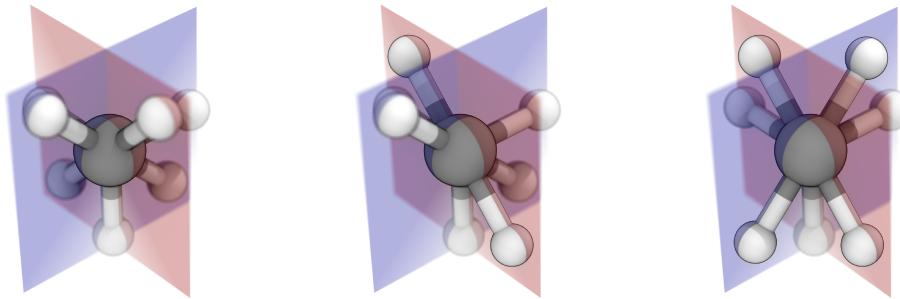

Figure S1: Three different structures with the same three-body representation

example of a group of three distinct structure having the same 3-body representation is shown in Fig. S1. The atomic coordinates associated with the three structures are:

$$\left(\begin{bmatrix}cos(0)\\sin(0)\\0\end{bmatrix}\begin{bmatrix}cos(-\frac{\pi}{4})\\sin(-\frac{\pi}{4})\\0\end{bmatrix}\begin{bmatrix}cos(-\frac{\pi}{2})\\sin(-\frac{\pi}{2})\\0\end{bmatrix}\begin{bmatrix}cos(\frac{3\pi}{4})\\sin(\frac{3\pi}{4})\\0\end{bmatrix}\begin{bmatrix}0\\cos(\frac{-3\pi}{4})\\sin(\frac{-3\pi}{4})\end{bmatrix}\begin{bmatrix}0\\cos(-\frac{\pi}{2})\\sin(-\frac{\pi}{2})\end{bmatrix}\begin{bmatrix}0\\cos(\frac{\pi}{4})\\sin(\frac{\pi}{4})\end{bmatrix}\right), \quad (29)$$

$$\left(\begin{bmatrix}cos(0)\\sin(0)\\0\end{bmatrix}\begin{bmatrix}cos(-\frac{\pi}{4})\\sin(-\frac{\pi}{4})\\0\end{bmatrix}\begin{bmatrix}cos(-\frac{\pi}{2})\\sin(-\frac{\pi}{2})\\0\end{bmatrix}\begin{bmatrix}cos(\frac{3\pi}{4})\\sin(\frac{3\pi}{4})\\0\end{bmatrix}\begin{bmatrix}0\\cos(\frac{3\pi}{4})\\sin(\frac{3\pi}{4})\end{bmatrix}\begin{bmatrix}0\\cos(-\frac{\pi}{2})\\sin(-\frac{\pi}{2})\end{bmatrix}\begin{bmatrix}0\\cos(-\frac{\pi}{4})\\sin(-\frac{\pi}{4})\end{bmatrix}\right), \quad (30)$$

$$\left(\begin{bmatrix}cos(0)\\sin(0)\\0\end{bmatrix}\begin{bmatrix}cos(\frac{\pi}{4})\\sin(\frac{\pi}{4})\\0\end{bmatrix}\begin{bmatrix}cos(-\frac{\pi}{2})\\sin(-\frac{\pi}{2})\\0\end{bmatrix}\begin{bmatrix}cos(-\frac{3\pi}{4})\\sin(-\frac{3\pi}{4})\\0\end{bmatrix}\begin{bmatrix}0\\cos(\frac{3\pi}{4})\\sin(\frac{3\pi}{4})\end{bmatrix}\begin{bmatrix}0\\cos(-\frac{\pi}{2})\\sin(-\frac{\pi}{2})\end{bmatrix}\begin{bmatrix}0\\cos(-\frac{\pi}{4})\\sin(-\frac{\pi}{4})\end{bmatrix}\right). \quad (31)$$

This example can be obtained by extending the counterexample discussed above, adding further points. Start from the planar example showed in Fig. 1a, and label the two degenerate configurations $\mathcal{X}^+$ and $\mathcal{X}^-$. These structures can be extended by adding an arbitrary number of points to both configurations, lying on a plane perpendicular to the plane of the figure, without lifting the degeneracy. Let's take structure $\mathcal{X}^+$, and extend it with the same structure $\mathcal{X}^+$ in the perpendicular plane, forming a configuration $\mathcal{X}^{++}$. Similarly, one can obtain two additional structures $\mathcal{X}^{--}$ and $\mathcal{X}^{+-}$. In all cases the set of distance-angle triplets consist of three groups. One that contains the neighbors from the initial structure, one that contains those from the additional atoms in the perpendicular plane, and a third one for which one neighbor comes from the initial structure and the other from the additional atoms. Since the structures $\mathcal{X}^+$ and $\mathcal{X}^-$ are indistinguishable we automatically see that first two distance-angle groups coincide in all the three extended structures. One only needs to check whether the 3-body correlations between the two groups, which are determined by the mutual placement of the structures on the two orthogonal planes, are the same in the three structures. The relative orientation is in turn determined by a single angle. We could determine, by trial-and-error, a value of the mutual orientation that leads to the three structures being degenerate with each other.

## 2 Bispectrum

Fig. S2 shows how to generate pairs of atom configurations with the same origin-centred bispectrum. In the first three panels an atom sits at the origin $O$ and at every polygon vertex $A$, $B$, $C$, etc. By adding another atom an arbitrary distance above [+] or below [−] the plane of the page, such that the lines $O[+]$ and $O[-]$ are orthogonal to the plane, one obtains two atom configurations (+) and (−) which share the same origin-centred bispectrum, even if they cannot be superimposed by proper rotations (they are chiral mirror images of each other).

In panel 1., the triangle $ABC$ lies in the plane of the page. Every tetrahedron containing $O$ and [+] as vertices can be superimposed on another tetrahedron containing $O$ and [−] as vertices by a rigid rotation. However, if two of the angles $a, b, c$ are different, the atom configurations are non-superimposable mirror images of each other. There are various free parameters: two of the angles, the circle radius, the distance separating [+] and [−] and even the number of atoms on the circle. This last freedom is demonstrated in panel 2. Panel 3. shows the simplest example we have found of two atom configurations which are not congruent – i.e. that cannot be superimposed by either proper or improper rotations – but share the same bispectrum. The congruent triangles $ABC$ and $DEF$ are parallel with the plane of the page and share the same distance to it. As with the other examples, every tetrahedron containing $O$ and [+] as vertices can be superimposed on another tetrahedron containing $O$ and [−] as vertices by a rigid rotation, but the atom configurations (+) and (−) are not congruent. The position of the points [+] and [−] are shown explicitly in panel 4, which corresponds to the same construction as panel 3 when viewed from the side. There are six free parameters: two of the angles in $a, b, c$, the torsion angle between the triangles, the distance between the triangles, the circle radius and the distance between [+] and [−]. This example can also be generalised as in panel 2. with any asymmetric polygon instead of a triangle.

In all of the examples depicted in the figure, one can add more atoms along the axis containing $O$, [+], [−] to the atom configuration (+), provided one also adds more atoms along the same axis to (−), such that every atom along this axis in (+) maps on to a corresponding atom in (−) under a reflection through the plane of the page.

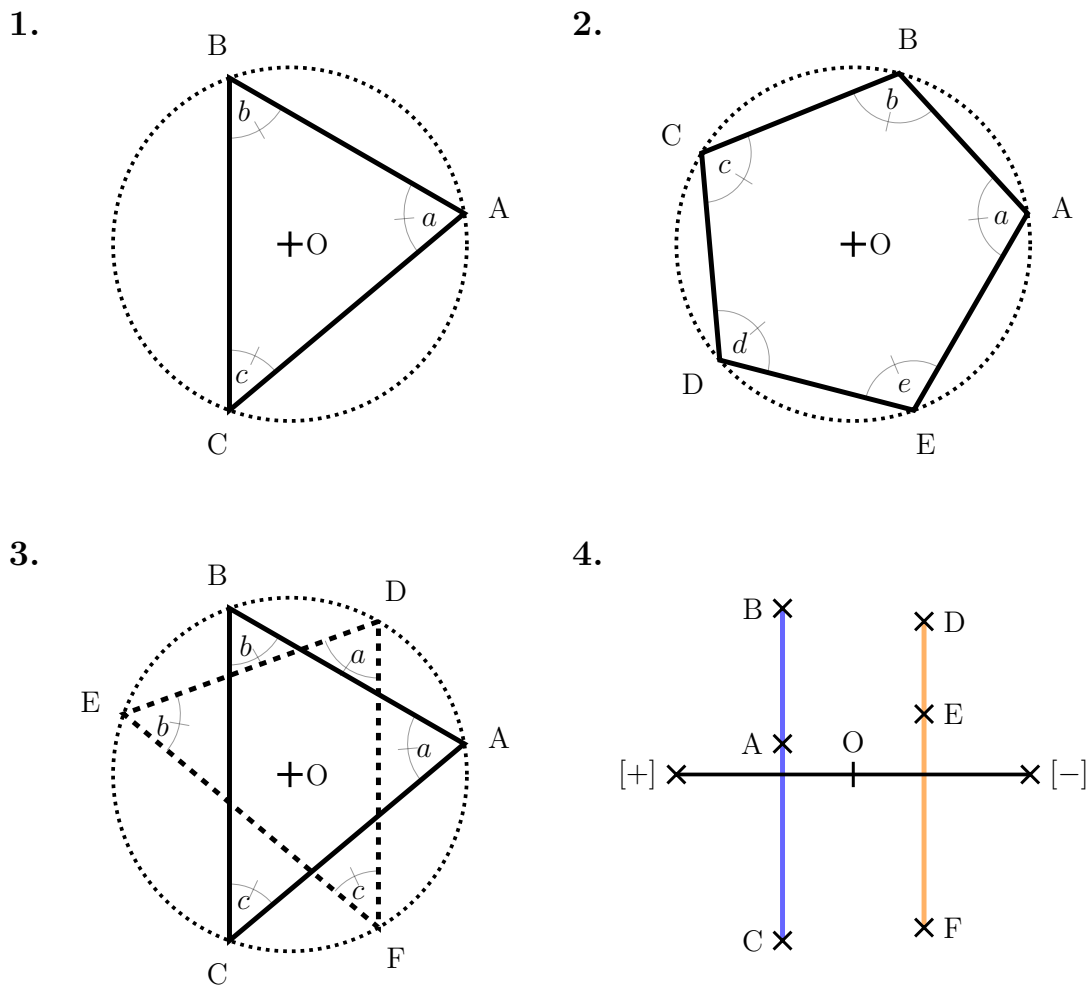

Figure S2: Four diagrams showing how to generate pairs of atom configurations with the same origin-centred bispectrum. In the first three panels an atom sits at the origin $O$ and at every polygon vertex $A$, $B$, $C$, etc. By adding another atom an arbitrary distance above [+] or below [−] the plane of the page, such that the lines $O[+]$ and $O[-]$ are orthogonal to the plane, one obtains two atom configurations (+) and (−) which share the same origin-centred bispectrum. Panel 4 shows the construction in panel 3 from the side.

# 3 Regression models

In order to verify how the discussed degeneracies influence the performance of machine learning potentials, we used Kernel Ridge Regression (KRR). The following atomic kernel was used,

$$k_{ij} = \exp\left(-(2 - 2\sum_q f_{iq}f_{jq})/\sigma_k^2\right), \tag{32}$$

where $f_{iq}$ and $f_{jq}$ are features (power spectrum or bispectrum) describing the environment of atoms $i$ and $j$ respectively. We used atom-centred Gaussians with a width of 0.5Å to construct the atomic density. Since all atoms are at the same distance from the centre, we set a cutoff of 2Å and computed SOAP features with $n_{\max} = 1$ Gaussian type orbital radial basis function, and $l_{\max} = 4$ angular channels. In the case of the C-centered models (top and middle panels of Fig. 3) this kernel was applied directly, while in the case of the C and H-centered models (bottom panel of Fig. 3) we used the standard modification

$$K_{IJ} = \sum_{ij} k_{ij}, \tag{33}$$

where the $K_{IJ}$ is the kernel between structures $I$ and $J$, the summation runs over all atoms in structures $I$ and $J$, and $k_{ij}$ is the atomic kernel in Eq. (32). The exponential kernel parameter $\sigma_k^2$ and the $L^2$ regularization parameter were optimized by cross-validation for every learning setup. The predictions plotted in Fig. 3 are obtained by 2-fold cross-validation, and averaged over 100 random shuffles of the dataset.

In the case of silicon we used $n_{\max} = 8$ radial basis functions and $l_{\max} = 6$ spherical harmonics channes. This led to an enormous number of bispectrum components. To avoid artefacts from the huge difference in dimensionality between the 3 and 4-body representations, we applied CUR feature selection to filter out 300 features from the powerspectrum and likewise for the bispectrum. We used the kernel introduced earlier in KRR with $\sigma_k^2 = 0.02$ for the short range model ($r_c = 2.8$ Å) and $\sigma_k^2 = 0.2$ for the long range model ($r_c = 5.0$ Å). The $L^2$ regularization parameter was chosen independently for every training setup using cross-validation.

# 4 Dimensionality of the degenerate set

We briefly summarize our theoretical findings on the co-dimension of the degenerate set for the 3-body descriptor.

First, it is immediate that for one or two neighbours the histogram of triangles determines the neighbourhood up to symmetries. For three neighbours the same statement is still true, and follows from the fact that a tetrahedron is uniquely defined by the three faces which are given by the three $1 \leq i < j \leq 3$ elements of the histrogram of triangles.

For four neighbours we have constructed a co-dimension 2 manifold of degenerate neighbourhood configurations. We now proceed to prove that this co-dimensionality is sharp: Given the histogram of triangles $\{(r_{ij}, r_{ij'}, \omega_{ijj'})\}$ we can define two polynomials

$$p_r = \prod_{j<j'}(r_j^2 - r_{j'}^2), \qquad p_n = p_D(\{\mathbf{r}_{ij}\}) \prod_{kk'' \neq jj'}(r_{kk'}^2 - r_{jj'}^2)$$

where $p_D$ is a polynomial defined in [1] with the following property (Theorem 2.6 in [1]): If $p_D(\{\mathbf{r}_j\}) \neq 0$ then the configuration $\{\mathbf{r}_j\}$ can be reconstructed from the histogram of distances $\{r_{jj'}\}$.

We now claim that, if either $p_r$ or $p_D$ is non-zero then the neighbourhood $\{\mathbf{r}_{ij}\}$ can be reconstructed up to symmetries. If $p_r$ is non-zero then all distances $r_{ij}$ are distinct hence the Gram matrix can be reconstructed and therefore also the neighbourhood (see the main text for more details).

If $p_r = 0$ but $p_n \neq 0$ (and hence $p_D \neq 0$) then according to [1, Thm. 2.6] we can reconstruct the positions of the neighbours $\mathbf{r}_{ij}$ up to an isometry and permutation which is a free symmetry, and up to an unknown *translation* which we now need to reconstruct. To this end we note that the added condition $r_{kk'} \neq r_{jj'}$ in the definition of $p_n$ means that the edges are all labelled. From these labels we can then get the labels for the corners. The issue that remains is that we still need to attach distances to the

corners: If all $r_{ij}$ are distinct, then there is nothing to prove. If, say $r_{i1} = r_{i2}$ then this means we can identify these two lengths from the triangle $(r_{i1}, r_{i2}, \omega_{i12})$. But knowing $r_{i1}$ we can identify $r_{ij}$ from $(r_{i1}, r_{ij}, \omega_{i1j})$. Therefore we now know all distances $r_{ij}$ to the corners $\mathbf{r}_{ij}$.

We can now return to determining the centre. If we have four or more neighbours that do not lie in a plane, then there is one possible centre. If all points lie in a plane then there are two possible centres but they are equivalent up to a reflection. If all points lie on a line then there is a circle of possible centres which are all equivalent up to rotations.

Thus, we have shown that for four or more neighbours, the set of degenerate neighbourhood structures are contained in an algebraic manifold which has at least co-dimension two.

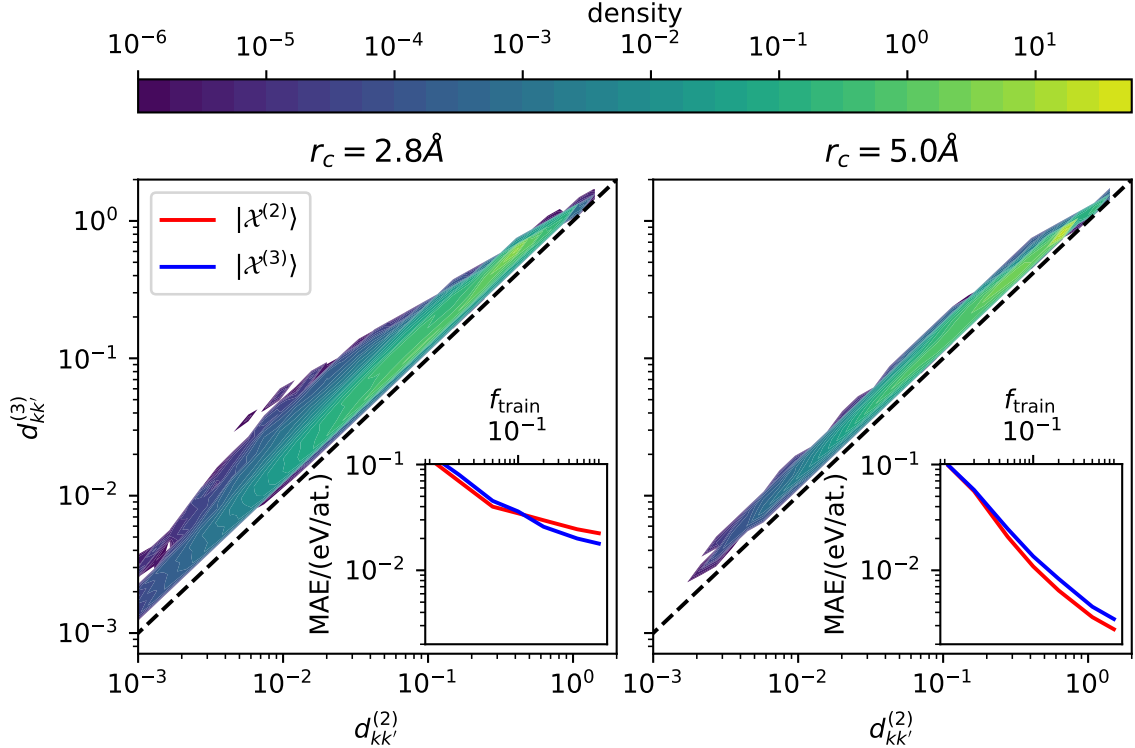

Figure S3: An analysis of atom-centered representations of a database of Si structures. The main panels show the joint distribution of environment distances, $d^{(2)}_{kk'}$ and $d^{(3)}_{kk'}$, computed respectively from $|\mathcal{X}^{(2)}\rangle$ and $|\mathcal{X}^{(3)}\rangle$. Insets show learning curves for a model predicting the energy of the structures, built using 3 and 4-body representations and a Gaussian kernel. Left panels are based on a cutoff of 2.8 Å, right panels are based on a cutoff of 5.0 Å.

## 5   An analysis of atom-centered representations of a database of Si structures

To assess the relevance of the considerations in the main text for more complex modelling problems, we consider the database of silicon structures from Ref. 2. We computed an expansion of $|\mathcal{X}^{(2)}\rangle$ and $|\mathcal{X}^{(3)}\rangle$ on a basis of radial functions and spherical harmonics. We also computed the distance $d^{(\nu)}_{ii'}$ between the environments centered on atoms $i, i'$, as the Euclidean distance between the corresponding feature vectors. To reduce the impact of the much larger dimensionality of $|\mathcal{X}^{(3)}\rangle$, we selected using CUR [3, 4] the 300 most significant components.

The left panel in Fig. S3 shows clear signals of the vicinity to a degenerate manifold, when using a short-range cutoff that only encompasses the nearest-neighbors shell. Indeed, pairs of environments can

be found for which $d^{(3)}_{ii'}$ is almost 30 times larger than $d^{(2)}_{ii'}$, similar to what we observed in the artificial CH$_4$ dataset (Fig. 2d). We propose that the joint distance distributions can be used as a diagnostic tool to identify the presence of (near)-degenerate pairs of structures in a dataset.

The right panel shows the same figure for a longer cutoff. Here, the resolving power of $|\mathcal{X}^{(2)}\rangle$ is comparable to that of the higher-order representation, with $d^{(2)}$ being at most a factor of 2 smaller than $d^{(3)}$. This suggests that degenerate structures become "less dense" as the number of neighbors is increased, which is consistent with the empirical observation that adding points to the construction in Fig. 1 increases the codimension of this particular set of indistinguishable configurations. At present, we cannot rigorously prove or disprove a more general statement.

These observations are reflected in the performance of a ML model for the cohesive energy of Si configurations. In the case of the short-range cutoff, the model based on $|\mathcal{X}^{(3)}\rangle$ marginally outperforms that based on $|\mathcal{X}^{(2)}\rangle$, while the opposite is true for the 5.0Å cutoff model. The differences are small, and sensitive to the details of the model. This suggests that when learning an atom-centred decomposition of an extensive property, such as the energy, at least to the relatively crude accuracy level of this example, the performance is minimally affected by the presence of near-degenerate structures. A low-body-order model may even outperform its higher-order counterpart, e.g., due to the linear link to 2 and 3-body potentials that provide the leading-order contribution to the energy. However, it is clear that, in addition to long-range interactions [5], the presence of degeneracies that we have discovered will also limit the ultimate accuracy that can be achieved, as we demonstrate in the main text in the case of a database of methane structures.

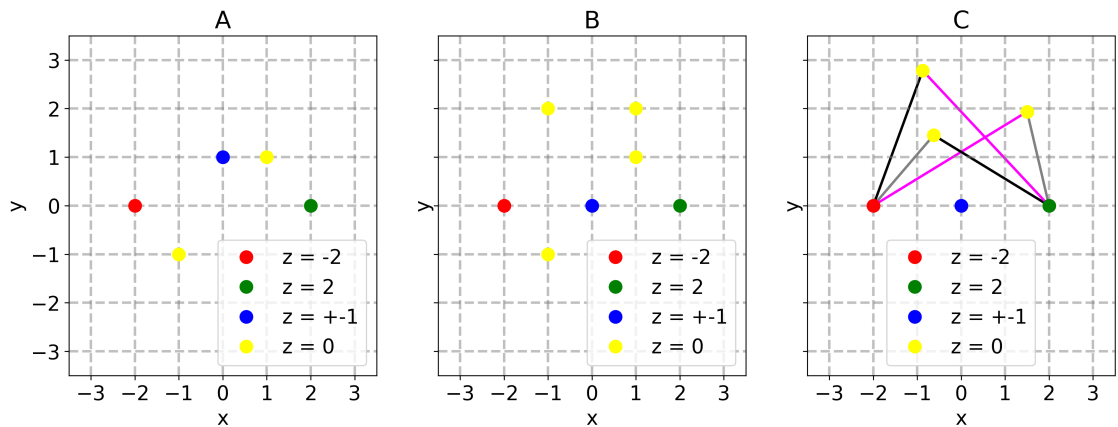

Figure S4: 2-body global counterexamples. Each panel schematically represents a pair of structures that are different, but cannot be differentiated knowing the lists of distances around each of the points. In case of C the segments with the same color have the same length

## 6 Counterexample for set of sets of distances

An example of two different structures which have the same set of sets of distances around every atom is given in panel A of Fig. S4.

Both structures have 5 atoms and differ only in the position of the blue one, which is placed at $z = +1$ in one and at $z = -1$ in the other . Let us show that the set of sets of distances for the two structures coincide. 1) The sets of distances for yellow atoms coincide in both structures. Indeed these atoms are placed in the $z = 0$ plane, and the position of the blue atom is symmetric around this plane. 2) The set of distances for the blue atom is the same in both structures, from the previously mentioned observation that the distances between blue and yellow atoms are the same, while the distances to red and green atoms are swapped. Indeed the distance between the blue and red atom in the positive structure is the same as the distance between the blue and green atom in the negative structure, and vice-versa. 3) The sets of distances for red and green atoms are swapped between two structures. It follows from the

fact that the green-yellow set of distances coincide with the red-yellow set of distances (each set has two entries) and from the previously mentioned swapping of blue-red and blue-green distances.

Using this logic one can understand that in order to construct a global 2-body counterexample one should place the yellow $z = 0$ points in such a way to 1) ensure the same blue-yellow and green-yellow sets of distances and 2) break both symmetries: point inversion around $x = 0, y = 0$ and inversion with respect to the $x = 0$ plane. (Here the projection of the blue point is also counted for breaking. The second condition is needed to ensure both structures are different with respect to rotations and other rigid motions.) This leads to a very rich high-dimensional manifold of global degeneracy. Other examples are shown in panels B and C of Fig. S4. Needless to say that $z$-levels can be chosen arbitrarily.

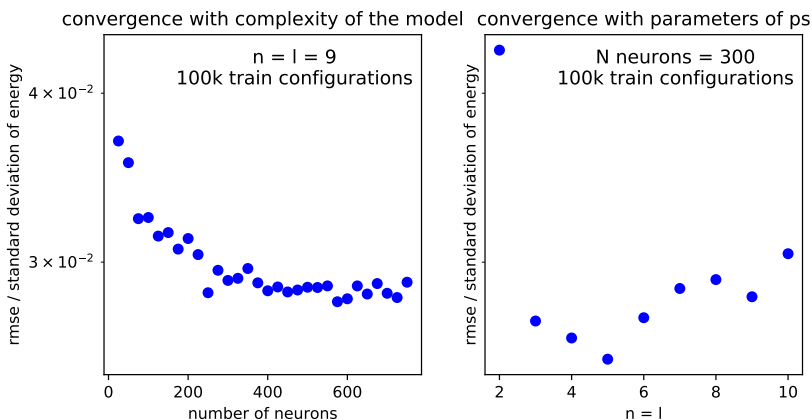

Figure S5: Dependence of the accuracy with machine learning model and power spectrum parameters. The analysis was performed with 100k training configurations due to the difficulties of fitting on more structures where power spectrum/bispectrum components do not fit into the RAM.

# 7  Random CH$_4$ models

The purpose of the following experiments is to numerically demonstrate that, due to degeneracy, a model based on 3-body features becomes much less accurate compared to one based on 4-body features provided one considers a sufficiently rich train set, and aims at achieving low prediction error. They also provide strong evidence that the accuracy saturates with increasing training set size at a sufficiently large number of training configurations. Since the effect of 3-body degeneracies is much more subtle than the effect of 2-body degeneracies, a huge amount of data is required to demonstrate it – particularly if structures are generated randomly, as we do here, to make sure that our conclusions are not biased by a particular choice of the data set. Thus, we calculated ab initio energies for 3 million random methane molecules. The molecules were generated as follow: First, H atoms were placed randomly (with a uniform distribution) within a cube centered on C atom. Second, molecules containing two atoms closer together than 0.5Å were discarded. Finally, we kept only those structures from the previous pool for which the self-consistent field method converged in a reasonable number of iterations (5 times more than the default value). Calculations were performed with Psi4 using GGA DFT and a single-zeta basis set.

As the machine learning model we decided to use neural networks as the universal functions approximators. We found that multilayer perceptrons (MLP) with two hidden layers performed best from a practical point of view (generalizing ability, speed of fitting and total number of parameters to achieve a given accuracy) compared to MLP with a different number of hidden layers. We always used the same number of neurons in both hidden layers, 300 in the final experiments shown in Fig. 3 (b). Since we are interested only in energies, we used the rectified linear unit (ReLU) activation function. In addition we added batch normalization to speed up convergence.

In all experiments we used the early stopping technique. The test data set (80k structures) which was used to evaluate the errors was the same for all experiments, including the linear regression models

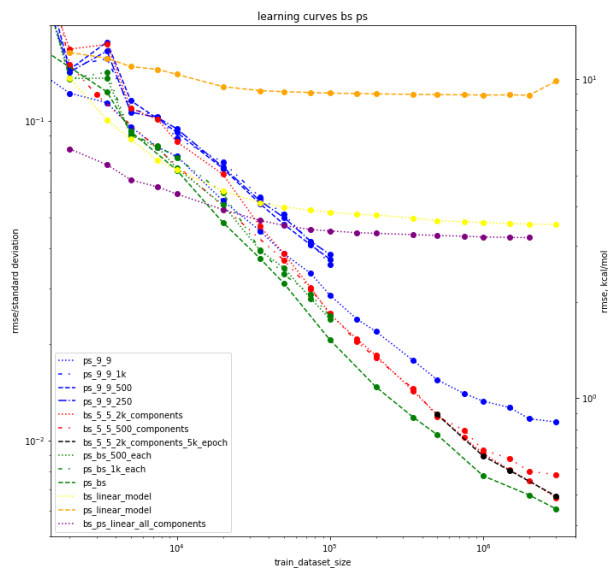

Figure S6: Full set of learning curves for the different models we evaluated.

we also computed for comparison. The Adam optimizer was used along with the ReduceLROnPlateau learning rate scheduler from PyTorch. The fitting steps were grouped into epochs, each of which consisted of passing 100k structures through the neural network, independent of the current data set size. In all experiments, the fitting procedure consisted of 2000 epochs, which we found to be enough for complete convergence with significant margin. As the 3-body features we used the SOAP power spectrum with convergence parameters $n_{max} = l_{max} = 9$. The cutoff radius was set in such a way to ensure that whole molecule was within the cutoff sphere for every atom. It is important to make sure that Fig. 3 (b) indicates the issues caused by degradation of the 3-body correlation itself, not the effect of insufficient $n_{max}$ or $l_{max}$, or insufficient flexibility of the machine learning model. It can be seen very clearly in Fig. S5 that the chosen parameters are sufficient with a good margin.

The convergence parameters of the bispectrum were $n_{max} = l_{max} = 5$. It appeared that the number of components was too high for fitting, and thus we used 2k PCA components. Fig. 3 (b) shows very clearly that with a sufficient amount of data there is significant slow down of the improvement of the power-spectrum model, in contrast to the bispectrum model. Moreover, one can even assume the saturation of the accuracy. We investigated the effect on the learning performance of changing the number of PCA components and using both power spectrum and bisectrum components as features in the same model. These learning curves are shown in Fig. S6 In this figure several trends can be distinguished. First, for small training set sizes, the models based on small numbers of PCA components perform the same or slightly better than the models based on large numbers of PCA components (that suffer from overfitting), but their learning rates are worse. Second, the model based on the 2- and 3-body correlation performs slightly better than the one based on only 3-body correlations. Also, the linear model based on all features is surprisingly accurate for small training set sizes.


}

\end{document}